\documentclass[a4paper,12pt]{article}

\newcommand{\sect}[1]{\setcounter{equation}{0}\section{#1}}

\begin{document}
\topmargin 0pt \oddsidemargin 0mm

\renewcommand{\thefootnote}{\fnsymbol{footnote}}
\begin{titlepage}
\begin{flushright}
KU-TP 013
\end{flushright}

\vspace{1mm}
\begin{center}
{\Large \bf Deconfinement Transition of AdS/QCD at ${\cal
O}(\alpha'^3)$ }

\vspace{10mm} {\large
Rong-Gen Cai\footnote{Email address: cairg@itp.ac.cn}}\\
\vspace{5mm}
{ \em Institute of Theoretical Physics, Chinese Academy of Sciences, \\
   P.O. Box 2735, Beijing 100080, China}\\
   \vspace*{1cm}
{\large Nobuyoshi Ohta\footnote{Email address:
ohtan@phys.kindai.ac.jp}} \\
\vspace{5mm} {\em Department of Physics, Kinki University,
Higashi-Osaka, Osaka 577-8502, Japan} \\

\end{center}
\vspace{5mm} \centerline{{\bf{Abstract}}} \vspace{5mm} We study
the confinement/deconfinement phase transition of holographic
AdS/QCD models by using Ricci flat $AdS_5$ black holes up to
${\cal O}(\alpha'^3)$, which corresponds to the $\lambda$
expansion correction in the dual field theory to $\lambda^{-3/2}$,
where $\lambda$ is the 't Hooft coupling constant. We consider two
cases: one is the hard-wall AdS/QCD model where a small radius
region of the $AdS_5$ is removed; the other is the case where one
of spatial coordinates for the $AdS_5$ space is compactified,
resulting in Witten's QCD model in $2+1$ dimensions. We find that
in the hard-wall AdS/QCD model, the deconfinement temperature
decreases when the $\lambda$ expansion corrections are taken into
account, while in Witten's QCD model, the deconfinement transition
always happens when the ratio of inverse temperature $\beta$ to
the period $\beta_s$ of the compactified coordinate decreases to
one, $\beta/\beta_s=1$, the same as the case without the ${\cal
O}(\alpha'^3)$ correction.

\end{titlepage}

\newpage
\renewcommand{\thefootnote}{\arabic{footnote}}
\setcounter{footnote}{0} \setcounter{page}{2}
\sect{Introduction}

The remarkable AdS/CFT correspondence~\cite{AdS} conjectures that
string/M theory in an anti-de Sitter space (AdS) times a compact
manifold is dual to a large $N$ strongly coupling conformal field
theory (CFT) residing on the boundary of the AdS space.  At finite
temperature, Witten~\cite{Witten} argued that the thermodynamics
of black holes in AdS space can be identified with that of the
dual strongly coupling field theory in the high temperature limit.
In addition, it is well known that there exists a phase transition
between the Schwarzschild-AdS black hole and thermal AdS space,
the so-called Hawking-Page phase transition~\cite{HP}: the black
hole phase dominates the partition function in the high
temperature limit, while the thermal AdS space dominates in the
low temperature limit. This phase transition is a first order one,
and is interpreted as the confinement/deconfinement phase
transition in the dual field theory~\cite{Witten}.

A special example of the AdS/CFT correspondence is that type IIB
string theory in $AdS_5\times S^5$ is dual to a four dimensional
${\cal N}=4$ supersymmetric Yang-Mills (SYM) theory on the
boundary of $AdS_5$. Acting as the near horizon geometry of
D3-branes, the $AdS_5$ space is the one in the Poincare
coordinates and dual field theory resides on a manifold with
topology $R\times R^3$. In that case, the dual field theory is
always in the deconfinement phase~\cite{Wilson,Wilson2}. In order
to realize a confined phase, the authors of \cite{PS2} proposed a
simple model, where a small radius region is removed from the
$AdS_5$ space in the Poincare coordinates without compact
directions. Note that the radial coordinate in the AdS space is
dual to an energy scale in the dual field theory. Therefore
removing a small radius region of AdS space corresponds to
introducing an IR cutoff and a mass gap in the dual theory. The
model in \cite{PS2} is called hard-wall AdS/QCD model in $3+1$
dimensions. Although the hard wall model is somewhat rough, it
turns out that the model can give some realistic, semiquantitative
description of low energy QCD~\cite{EKSS,DP}.

Recently, Herzog~\cite{Herzog} has shown that in this simple hard
wall model of AdS/QCD, the confinement/deconfinement phase
transition occurs via a first order Hawking-Page phase transition
between the low temperature thermal AdS space and high temperature
AdS black hole in the Poincare coordinates. This deconfinement phase
transition has been studied in various
cases~\cite{BBF}-\cite{KLNP}. For instance, in \cite{CS} the
deconfinement transition has been studied for the AdS/QCD  model
in curved spaces and with chemical potential; the authors of
\cite{KLNP} have discussed the effect of matter on the
deconfinement temperature.

Another scenario to realize the confinement of gauge field was
proposed by Witten~\cite{Witten}: Consider one of spatial directions
on the world volume of D3-branes is compactified, and fermions along
that direction are anti-periodic so that supersymmetry is broken. In
that case, the near horizon geometry $AdS_5\times S^5$ can be
regarded as gravity dual of low-energy QCD model in $2+1$
dimensions. When one of spatial directions is compactified, Horowitz
and Myers~\cite{HM} have shown that there is a lower mass solution
than the pure AdS space. This solution is called AdS soliton. Viewed
AdS soliton as a reference background, it was shown that there is a
Hawking-Page phase transition between Ricci flat AdS black holes and
AdS soliton~\cite{SSW}-\cite{BD}. It is found that the
deconfinement transition is determined by the ratio of inverse
Hawking temperature of the Ricci flat black hole to the period of
the compactified direction.

In this note we will discuss the deconfinement transition of the
above two holographic AdS/QCD models. Specially we pay attention on
the effect of the terms ${\cal O}(\alpha'^3)$ on the deconfinement
temperature. In type IIB supergravity, the leading correction of
$\alpha'$ expansion to the low-energy effective action is of the
form $\alpha'^3 R^4$. The $\alpha'$ expansion
is dual to $\lambda$ expansion in the dual field theory,
here $\lambda$ is the 't Hooft coupling constant of gauge field.
Thus the term ${\cal O}(\alpha'^3)$ in the supergravity action
corresponds to the correction $\lambda^{-3/2}$ in the QCD models.

The organization of this paper is as follows. In the next section we
first review the $AdS_5$ black hole solution, up to the correction
${\cal O}(\alpha'^3)$. In Sec.~III we discuss the deconfinement
phase transition of the hard-wall AdS/QCD model. Sec.~IV is devoted
to the case of Witten's QCD model in $2+1$ dimensions. We end the
paper with conclusion and discussion in Sec.~V.

\sect{$AdS_5$ black holes at ${\cal O}(\alpha'^3)$}

The near horizon geometry of black D3-branes can  be described by a
Ricci flat $AdS_5$ black hole times a constant radius $S^5$
\begin{equation}
\label{2eq1}
ds^2=\frac{r^2}{L^2} (-fdt^2 + dx_1^2+dx_2^2+dx_3^2) +
\frac{L^2}{r^2} f^{-1}dr^2 +L^2 d\Omega_5^2,
\end{equation}
where $f=1-r_0^4/r^4$,  $L^4= g_{\rm YM}^2N \alpha'^2$,  $N$ is the
number of D3-branes and is the rank of dual gauge field with SYM
coupling constant $g^2_{\rm YM}=4\pi g_s$, $\lambda =g_{\rm YM}^2N $
is the 't Hooft coupling constant for the $SU(N)$ gauge
theory~\footnote{Here the notations of the SYM coupling $g_{\rm
YM}^2$ and the 't Hooft coupling constant are different from those
in \cite{SKT} by a factor $2$, for more details see \cite{Sin}.},
$r_0$ is the mass parameter of the solution.

The black hole horizon is located at $r=r_0$.
The associated inverse Hawking temperature of the black hole solution (\ref{2eq1}) is
\begin{equation}
\beta_0= \pi L^2/r_0.
\end{equation}
 The 5-dimensional effective action obtained from
10-dimensional type IIB supergravity compactifying on $S^5$ is
\begin{equation}
\label{2eq2}
I_0 =-\frac{1}{16\pi G_5} \int d^5x
\sqrt{g_5}\left (R_5+\frac{12}{L^2}\right).
\end{equation}
Calculating the action yields the free energy and entropy of the
black hole~\cite{SKT}
\begin{equation}
\label{2eq4}  F_0= -\frac{\pi^2}{8}N^2V_3T_0^4, \ \ \ S_0=
\frac{\pi^2}{2}N^2 V_3 T^3_0
\end{equation}
where $V_3$ is the volume of space spanned by $x_1$, $x_2$ and
$x_3$.  On the other hand, the entropy of the weak coupling limit of
the $SU(N)$ SYM theory, where it reduces to that of $8N^2$ free
massless bosons and fermions, is
\begin{equation}
\label{2eq6}
S_{\rm YM}= \frac{2\pi^2}{3} N^2V_3 T^3_0.
\end{equation}
This means that $S_0= \frac{3}{4}S_{\rm YM}$. Note that according to
the dictionary of AdS/CFT correspondence, the black hole entropy is
equivalent to that of SYM theory in the limit of $N \rightarrow
\infty$ and large 't Hooft coupling constant $\lambda$. The
relations $L^4/l^4_p \sim N$ and $L^4/l_s^4 \sim g^2_{\rm YM}N$ tell
us that only in that limit both the $\alpha'$ and loop corrections
to the solution (\ref{2eq1}) can be neglected.

In \cite{SKT}, the $\alpha'$ corrections have been considered in
the dual gravity side. In type IIB supergravity, the leading
correction terms are of the form $\alpha'^3 R^4$. The tree level
type IIB string effective action has the following form
\begin{equation}
\label{2eq7}
I = -\frac{1}{16\pi G_{10}}\int d^{10}x\sqrt{g}\left(
R-\frac{1}{2}(\partial \phi)^2 -\frac{1}{4 \cdot 5!}(F_5)^2
+\cdots +\gamma e^{-3\phi/2}W+\cdots\right),
\end{equation}
where $\gamma = \frac{1}{8}\zeta(3)(\alpha')^3$, $F_5$ is assumed
self-dual, dots stand for other terms depending on antisymmetric
tensor field strengths and derivatives of dilaton, and
\begin{eqnarray}
 W &=& R^{hmnk}R_{pmnq}R_h^{\ rsp}R^q_{\ rsk}
 +\frac{1}{2}R^{hkmn}R_{pqmn}R_h^{rsp}R^q_{\ rsk}  \nonumber \\
   && +{\rm terms\ depending\ on\ the\ Ricci\ tensor}.
   \nonumber
\end{eqnarray}
 The field redefinition ambiguity allows one to change the coefficients
of terms involving the Ricci tensor (in essence, ignoring other
fields, one may use $R_{mn}=0$ to simplify the structure of $W$ as
the graviton legs in the 4-point string amplitude are
on-shell)~\cite{SKT}. Thus one can find a scheme where $W$ in
(\ref{2eq7}) depends only on the Weyl tensor
\begin{equation}
\label{2eq8}
W = C^{hmnk}C_{pmnq}C_h^{\ rsp}C^q_{\ rsk} +\frac{1}{2}
C^{hkmn}C_{pqmn}C_h^{\ rsp}C^q_{\ rsk}.
\end{equation}
The form of $W$ is special in the sense that the $AdS_5\times S^5$
is still the solution of the action (\ref{2eq7}) with self-dual
$F_5$ and a constant dilaton. The corrected free energy turns out to
be of the form
\begin{equation}
\label{2eq10}
F=F_0+\delta F = -\frac{\pi^2}{8} N^2 V_3 T_0^4 \left( 1+\frac{15}{8}
\zeta (3) \lambda ^{-3/2} \right ),
\end{equation}
and the corrected entropy of the black hole
\begin{equation}
\label{2eq11} S= \frac{\pi^2}{2}N^2V_3T^3_0 (1+\frac{15}{8}\zeta(3)
\lambda^{-3/2}).
\end{equation}
Indeed one can see that the leading correction is positive. If we
write the free energy in the following form
\begin{equation}
\label{2eq12} F= -f(\lambda ) \frac{\pi^2}{6}N^2V_3 T^4_0,
\end{equation}
it is expected that the function $f(\lambda)$ approaches $1$ for
small coupling $\lambda $, then for large coupling, one
has~\cite{SKT}
\begin{equation}
\label{2eq13}
f(\lambda) = \frac{3}{4} + \frac{45}{32}\zeta (3) \lambda^{-3/2}+ \cdots.
\end{equation}
 The $R^4$ correction to the AdS black holes
metric and  their thermodynamics have also been discussed in various
cases~\cite{PT}-\cite{Buch}. The corrections from the $R^4$ terms in
the noncommutative SYM theory have been discussed in \cite{CO}.

The correction to the $AdS_5$ black hole metric due to the $R^4$ term
in the action (\ref{2eq7}) can be obtained by following
\cite{SKT}.  We consider the following metric ansatz for the
10-dimensional metric in the Einstein frame,
\begin{equation}
\label{2eq15}
ds^2_{10} = e^{-\frac{10}{3} \nu(x)}g_{5mn}dx^mdx^n + e^{2\nu(x)}
d\Omega_5^2,
\end{equation}
where we have set $L=1$. Thus in what follows, $\gamma = \frac{1}{8}\zeta (3)
\lambda^{-3/2}$. Considering the standard ansatz for the
field strength and compactifying on $S^5$, one can obtain the
5-dimensional effective action
\begin{equation}
\label{2eq16} I_5= -\frac{1}{16 \pi G_5} \int d^5x \sqrt{g_5} \left[
R_5 -\frac{1}{2} (\partial \phi)^2 -\frac{40}{3}(\partial \nu)^2
-V(\nu)+\gamma e^{10\nu-\frac{3}{2}\phi}\left (W +{\cal
O}((\partial\nu)^2)\right)\right],
\end{equation}
where
$$ V(\nu)=8 e^{-\frac{40}{3}\nu }-20 e^{-\frac{16}{3}\nu }.$$ Up to
the  order ${\cal O(\gamma)}$, one has
\begin{equation}
\nu (r)= \frac{15\gamma}{32}\frac{r_0^8}{r^8}
\left(1+\frac{r_0^4}{r^4}\right) +{\cal O}(\gamma^2),
\end{equation}
while the dilaton field $\phi_1=\phi-\phi_0$ is
\begin{equation}
\phi_1= -\frac{45}{8}\gamma
\left(\frac{r_0^4}{r^4}+\frac{r_0^8}{2r^8}+\frac{r_0^{12}}{3r^{12}}\right)
+{\cal O}(\gamma^2).
\end{equation}
The 5-dimensional metric is given by~\cite{SKT}
\begin{equation}
\label{2eq19}
ds^2_5= H^2 (K^2d\tau^2 +P^2dr^2 +dx_1^2
+dx_2^2+dx_3^2),
\end{equation}
with
\begin{equation}
\label{2eq20}
H=r, \ \ \  K=e^{a+4b}, \ \ \ P=e^b,
\end{equation}
and
\begin{eqnarray}
\label{2eq21}
&& a =-2\ln r +\frac{5}{2} \ln (r^4-r_0^4)
-\frac{15}{2}\gamma \left( 25\frac{r_0^4}{r^4} +25
\frac{r_0^8}{r^8}-79\frac{r_0^{12}}{r^{12}}\right) +{\cal
O}(\gamma^2), \nonumber \\
&& b= -\frac{1}{2} \ln (r^4-r_0^4) +\frac{15}{2}\gamma
\left(5\frac{r_0^4}{r^4}+5\frac{r_0^8}{r^8}-19\frac{r_0^{12}}{r^{12}}\right)
+{\cal O }(\gamma^2).
\end{eqnarray}
Note that the leading order for both of the dilaton field $\phi$
and scalar $\nu$ is ${\cal O}(\gamma)$. Therefore the action
(\ref{2eq16}) can be reduced to
\begin{equation}
\label{2eq22} I_5= -\frac{1}{16 \pi G_5} \int d^5x \sqrt{g_5} (R_5
+12 +\gamma W),
\end{equation}
up to the leading order correction of the $\gamma$ term. Namely in
this order the dilaton field and scalar field $\nu$ have no
contribution to the Euclidean action and free energy. For the
black hole solution (\ref{2eq19}), the Hawking temperature is
\begin{equation}
\label{2eq23}
T \equiv 1/\beta = \frac{r_0}{\pi}(1+15\gamma) +{\cal O}(\gamma^2).
\end{equation}

\sect{Deconfinement transition in the hard-wall AdS/QCD}

Let us notice that even with the $R^4$ correction, the free energy
(\ref{2eq10}) is always negative, which means that the dual field
theory is in the deconfinement phase. It is true even in the
zero-temperature phase, which can be verified by calculating the
quark-anti-quark potential in the dual gravity
configuration~\cite{Wilson,Wilson2}. Recently it has been found
that one can realize a confined phase of gauge theory in the dual
$AdS_5$ gravity configuration by simply removing a small radius
region of $AdS_5$ spacetime~\cite{PS2}, which is equivalent to
introducing an IR cutoff in the dual field theory. The $AdS_5$
spacetime with an IR cutoff is called  hard-wall AdS/QCD model.
More recently  it has been found that the deconfinement phase
transition of the hard-wall AdS/QCD model can be realized via a
first-order Hawking-Page phase transition between a thermal AdS
space and a Ricci flat Schwarzschild-$AdS_5$ black
hole~\cite{Herzog}. In the following, we will consider the effect
of the $R^4$ term on the deconfinement phase transition.

Now we calculate the Euclidean action associated with the black
hole (\ref{2eq19}).
Due to an infinite volume, the action
(\ref{2eq22}) is divergent. To get a finite result, one has to
regularize the action by subtracting the contribution of a suitable
reference background. Here the suitable reference background is
obviously the one (\ref{2eq19}) with $r_0=0$. Namely we take the
$AdS_5$ spacetime in the Poincare coordinates as the reference
background. To finish this calculation, we introduce an UV boundary
at $r=R$ ($ R\gg r_0$) (Note that in the coordinates (\ref{2eq19}),
the radial coordinate $r$ is equivalent to an energy scale in the dual
field theory). On the other hand, we introduce an IR cutoff at
$r=x_0$ in order to realize the confinement in the hard-wall AdS/QCD
model. In the black hole
phase, the IR cutoff is taken as $r_{\rm max}=\max (r_0,x_0)$.
Therefore, one has
\begin{equation}
\label{2eq24} I_{5BH}= -\frac{N^2}{8\pi^2}V_3\beta \int^{R}_{r_{\rm
max}} dr \sqrt{g_5} (R_5 +12 +\gamma W).
\end{equation}
For the reference background, in order that the black hole solution
can be embedded into the reference background, the Euclidean time
for the background has to satisfy the following condition at the UV boundary
\begin{equation}
\beta' \sqrt{g_{00}(r_0=0, r=R)}= \beta \sqrt{g_{00}(r_0, r=R)},
\end{equation}
which gives
\begin{equation}
\label{2eq25}
\beta'= \beta \left(1-\frac{1}{2}(1+75\gamma)\frac{r_0^4}{R^4}
+{\cal O}(\frac{r_0^8}{R^8}) \right).
\end{equation}
Thus the contribution from the background is
\begin{eqnarray}
\label{2eq26}
I_{5REF} &=&-\frac{N^2}{8\pi^2}V_3\beta' \int^{R}_{x_0} dr
\sqrt{g_5} (R_5 +12 +\gamma W) \nonumber\\
&=& \frac{N^2}{4\pi^2} V_3 \beta' (R^4-x_0^4) \nonumber \\
&=& \frac{N^2}{4\pi^2}V_3\beta \left( R^4 -\frac{1}{2} (1+75\gamma)r_0^4-x_0^4\right).
\end{eqnarray}
Note that the $R^4$ term ($\gamma W$) gives no contribution to the action of
the reference background. The Euclidean action for the black hole solution is
\begin{eqnarray}
\label{2eq27}
I_{5BH} &=& -\frac{N^2}{8\pi^2}V_3\beta \int^{R}_{r_{\rm max}} dr
\left( -8r^3 +\gamma \left(\frac{360 r_0^{16}}{r^{13}} +\frac{960
r_0^{12}}{r^9}\right) +{\cal O}(\gamma^2)\right) \nonumber\\
&=& \frac{N^2}{4\pi^2}V_3\beta \left(R^4-r^4_{\rm max} +15r_0^{12} \gamma
\left(r_0^4(\frac{1}{R^{12}}-\frac{1}{r^{12}_{\rm max}}) + 4
(\frac{1}{R^8}-\frac{1}{r^8_{\rm max}})\right)\right).
\end{eqnarray}
Subtracting the contribution of the reference background from
(\ref{2eq27}) and taking $R\rightarrow \infty$, one obtains the
Euclidean action of the black hole
\begin{equation}
\label{2eq28}
{\cal I} = \frac{N^2}{4\pi^2} V_3\beta \left(-r^4_{\rm max}
+\frac{1}{2} (1+75\gamma)r_0^4+x_0^4 - 15r_0^{12} \gamma
\left(\frac{r_0^4}{r^{12}_{\rm max}} +\frac{4}{r^8_{\rm max}}\right)\right).
\end{equation}
Now it is the position to discuss the action.

\subsection{No IR cutoff}

Let us first consider the case without the IR cutoff, namely
$x_0=0$ and $r_{\rm max}=r_0$. In this case, one has
\begin{equation}
\label{2eq29} {\cal I} = -\frac{N^2}{8\pi^2} V_3\beta r_0^4 (1+75
\gamma)=-\frac{N^2\pi^2}{8}V_3T^3 (1+15\gamma).
\end{equation}
This is exactly the result obtained in \cite{SKT}, and this gives
the free energy in (\ref{2eq10}). Therefore no Hawking-Page phase
transition will happen in  this case.

\subsection{No higher order term}

In the case without the $R^4$ term, namely $\gamma =0$, one has
\begin{equation}
\label{2eq30}
{\cal I} =-\frac{N^2}{8\pi^2} V_3\beta \left(2r^4_{\rm max}-r_0^4-2 x_0^4 \right),
\end{equation}
which reproduces the results in \cite{Herzog,CS}: when
$x_0>r_0$, one has $r_{\rm max}=x_0$; the action is always
positive, and no Hawking-Page transition will occur. In contrast,
a Hawking-Page transition can happen when $x_0 <r_0$. In the latter
case, one has $r_{\rm max}= r_0$; the action is negative for $r_0^4>2x_0^4$ and
positive for $r_0^4<2x_0^4$; the action has the form
\begin{equation}
\label{in1}
{\cal I}=-\frac{N^2}{8\pi^2} V_3\beta \left(r_0^4-2
x_0^4 \right).
\end{equation}
The deconfinement temperature is
\begin{equation}
\label{in2}
T_c= 2^{\frac{1}{4}}\frac{x_0}{\pi}.
\end{equation}
This is precisely the result in \cite{Herzog}. Note that here
$x_0=1/z_0$ in \cite{Herzog}. Using the lightest $\rho$ meson mass,
one may conclude $z_0=1/(323 \mbox{ MeV})$. Note the fact that the
value $z_0$ is obtained by using the Bessel function in $AdS_5$
space~\cite{Herzog}, and the $W$ term does not change the geometry
of $AdS_5$. We conclude that the correction term $W$ will not give
rise to any change of the value $z_0$.

\subsection{Effects of IR cutoff and higher order term}

Now we consider the case where both the IR cutoff and the
$R^4$ term are present. Here we have two subcases:

1)  If the IR cutoff $x_0>r_0$, then one has $r_{\rm max}=x_0$,
and the action reduces to
\begin{eqnarray}
\label{2eq31}
{\cal I} &=&\frac{N^2}{8\pi^2} V_3\beta r_0^4 \left( 1+75\gamma
 - 30 \gamma
\left(\frac{r_0^{12}}{x^{12}_0} +\frac{4
r_0^{8}}{x^8_0}\right)\right) \nonumber \\
&=& \frac{N^2\pi^2}{8} V_3 T^3
\left (1 +15\gamma-30\gamma\left(\frac{T^{12}}{T^{12}_{IR}}
+4\frac{T^8}{T_{IR}^8}\right ) \right),
\end{eqnarray}
where we have defined $T_{IR}= x_0/\pi $. Although $\gamma<1$ and
$T/T_{IR}<1$, at first sight it appears that the action
(\ref{2eq31}) can change the sign; if so, it would imply that a
Hawking-Page phase transition would take place. However, this
conclusion is unreliable because in getting the action
(\ref{2eq31}), we have treated the terms concerning $\gamma$ as
perturbative terms, which indicates that the action is valid only
when the second and third terms are much less than the first term
in (\ref{2eq31}).  In that sense, the action (\ref{2eq31}) is
always positive, and the thermal $AdS_5$ space with an IR cutoff
dominates in the lower temperature phase, the dual field theory is
always in the confined phase. What we can conclude here is that
the action gets negative contribution from the higher order
effect.

2) If $x_0<r_0$, one has $r_{\rm max}=r_0$. In that case, the action
becomes
\begin{eqnarray}
{\cal I} &=& -\frac{N^2}{8\pi^2}V_3 \beta \left( (1+75\gamma)r_0^4 -2
x_0^4\right) \nonumber \\
&=& -\frac{N^2\pi^2}{8}V_3 T^3\left (1+15\gamma
 -\frac{2x_0^4}{\pi^4 T^4}\right ).
\end{eqnarray}
The action can change its sign. A Hawking-Page phase transition
happens when ${\cal I}=0$. In the high temperature phase, the
action is negative and dual field theory is in the deconfinement
phase, while it is in the confined phase in the low temperature
case. Clearly the phase transition temperature is
\begin{equation}
\label{2eq34} T_c= \frac{2^{1/4} x_0}{\pi} (1-\frac{15}{4}\gamma).
\end{equation}
When $\gamma=0$, it reduces to the one (\ref{in2}). We see from
(\ref{2eq34}) that the deconfinement temperature decreases if we
incorporate the correction ${\cal O}(\alpha'^3)$.

The energy of the dual field theory is
\begin{equation}
\label{2eq35} E \equiv \frac{\partial {\cal I}}{\partial \beta}
=\frac{3N^2\pi^2}{8}V_3 T^4\left (1+15\gamma
 +
\frac{2x_0^4}{3\pi^4 T^4}\right ),
\end{equation}
while its entropy is
\begin{equation}
\label{2eq36}
S =\frac{N^2\pi^2}{2}V_3 T^3(1+15\gamma).
\end{equation}
We see from (\ref{2eq35}) that indeed introducing an IR cutoff is
equivalent to a mass gap; the energy does not vanish even when the
temperature approaches zero, while the entropy is still the
same as the case without the IR cutoff.

\sect{Deconfinement  transition in Witten's QCD Model}

In this section we discuss the deconfinement transition in
Witten's QCD model in $2+1$ dimensions. To this aim, we have to
first get the corresponding soliton solution, which will act as
the reference background. The deformed AdS soliton at the order
${\cal O}(\alpha'^3)$ can be obtained by continuing the Euclidean
black hole solution (\ref{2eq19}) in the way $x_1 \rightarrow it$
\begin{equation}
\label{3eq1}
ds^2_5 = H^2_s (K^2_s d\tau^2 +P^2_sdr^2 -dt^2 +dx_2^2+dx_3^2),
\end{equation}
where $H_s$, $K_s$ and $P_s$ are still given by $H$, $K$ and $P$ in (\ref{2eq20}),
but replacing $r_0$ by a new constant $r_s$ in (\ref{2eq21}).
Now the coordinate $\tau$ is a spatial one. In order to remove the
conical singularity in the plane spanned by $\tau$ and $r$,
the  coordinate $\tau$ has to be a periodic one with period
\begin{equation}
\label{3eq2}
\beta_s= (1-15\gamma) \frac{\pi}{r_s} +{\cal O}(\gamma^2).
\end{equation}
Let us next consider the black hole solution in (\ref{2eq19})
\begin{equation}
\label{3eq3}
ds^2_5 = H^2 (-K^2 dt^2 +P^2dr^2 + dx_1^2 +dx_2^2+dx_3^2),
\end{equation}
with a compactified coordinate $x_1$ with period $\eta$. The dual field theory
resides on a manifold with topology $R^1 \times S^1\times R^2$, where $R^1$ stands for
time, $S^1$ represents the coordinate $x_1$ and $R^2$ is for
the coordinates $x_2$ and $x_3$. In this
case, a natural reference background is just the one (\ref{3eq3})
with $r_0=0$, namely the $AdS_5$ space in the Poincare
coordinates, but with a compactified coordinate $x_1$, like the
situation discussed in the previous section. However, as shown in
\cite{HM}, the AdS soliton  has a less energy than the $AdS_5$
space. In our case, the corresponding AdS soliton solution is
the one given by (\ref{3eq1}), up to ${\cal O}(\alpha'^3)$.
Therefore we will study the Euclidean action of the black hole
(\ref{3eq3}) by regarding the AdS soliton (\ref{3eq1}) as the
reference background. In order that the black hole solution can
be embedded into the reference background, at the UV boundary
$r=R$ ($R\gg r_0$ and $R\gg r_s$), we must have the matching conditions
\begin{equation}
\label{3eq4}
\beta HK(r=R) =\beta_b H_s(r=R), \ \ \ \eta H(r=R)=\beta_s
H_sK_s(r=R),
\end{equation}
where $\beta$ is the inverse temperature of the black hole
(\ref{3eq3}), which is still given by (\ref{2eq23}), while $\beta_b$ is
the period of the Euclidean time for the soliton solution
(\ref{3eq1}). Thus we have
\begin{eqnarray}
\label{3eq5}
\beta_b &=& \beta \left(1-\frac{1}{2}(1+75\gamma)\frac{r_0^4}{R^4}
+{\cal O}(\frac{r_0^8}{R^8}) \right), \nonumber \\
\eta &=&  \beta_s \left(1-\frac{1}{2}(1+75\gamma)\frac{r_s^4}{R^4}
+{\cal O}(\frac{r_s^8}{R^8}) \right).
\end{eqnarray}
The Euclidean action for the black hole is
\begin{eqnarray}
\label{3eq6}
I_{5BH} &=& -\frac{N^2}{8\pi^2}V_2 \eta \beta \int^{R}_{r_0} dr
\left( -8r^3 +\gamma \left(\frac{360 r_0^{16}}{r^{13}} +\frac{960
r_0^{12}}{r^9}\right) +{\cal O}(\gamma^2)\right) \nonumber\\
&=& \frac{N^2}{4\pi^2}V_2\beta \eta \left(R^4-r^4_0 +15r_0^{12} \gamma
\left(r_0^4(\frac{1}{R^{12}}-\frac{1}{r^{12}_0}) + 4
(\frac{1}{R^8}-\frac{1}{r^8_0})\right)\right) \nonumber \\
&=& \frac{N^2}{4\pi^2}V_2\beta \beta_s \left ( R^4 -(1+75\gamma)
r_0^4 -\frac{1}{2}(1+75 \gamma)r_s^4\right),
\end{eqnarray}
where $V_2$ is the volume for the space spanned by coordinates
$x_2$ and $x_3$ and in the last equality we have dropped terms
disappearing in the large $R$ limit. On the other hand, the
Euclidean action for the soliton is
\begin{eqnarray}
\label{3eq7} I_{5REF} &=& -\frac{N^2}{8\pi^2}V_2 \beta_b \beta_s
\int^{R}_{r_s} dr
\left( -8r^3 +\gamma \left(\frac{360 r_0^{16}}{r^{13}} +\frac{960
r_0^{12}}{r^9}\right) +{\cal O}(\gamma^2)\right) \nonumber\\
&=& \frac{N^2}{4\pi^2}V_2\beta_b \beta_s \left(R^4-r^4_s
+15r_s^{12} \gamma
\left(r_s^4(\frac{1}{R^{12}}-\frac{1}{r^{12}_s}) + 4
(\frac{1}{R^8}-\frac{1}{r^8_s})\right)\right) \nonumber \\
&=& \frac{N^2}{4\pi^2}V_2\beta \beta_s \left ( R^4 -(1+75\gamma)
  r_s^4 -\frac{1}{2}(1+75 \gamma)r_0^4\right).
\end{eqnarray}
In the last equality we have considered the large $R$ limit, once again.
The difference between the Euclidean actions for the black hole and soliton is
\begin{eqnarray}
\label{3eq8}
{\cal I} &=& -\frac{N^2}{8\pi^2}V_2\beta \beta_s
(1+75\gamma)(r_0^4-r_s^4) \nonumber \\
&=&-\frac{N^2 \pi^2}{8}V_2 \beta^{-3} \beta_s
(1+15\gamma)\left(1-\frac{\beta^4}{\beta_s^4}\right).
\end{eqnarray}
The energy of the black hole is, via $E=\partial {\cal I}/\partial
\beta$,
\begin{equation}
\label{3eq9}
E=\frac{N^2 \pi^2}{8}V_2\beta^{-4}\beta_s (1+15\gamma)
(3+\frac{\beta^4}{\beta_s^4}),
\end{equation}
and the entropy associated with the black hole is
\begin{equation}
\label{3eq10}
S= \frac{N^2 \pi^2}{2}V_2\beta^{-3}\beta_s (1+15\gamma).
\end{equation}
We see from the action ({\ref{3eq8}) that the action changes its
sign at $\beta=\beta_s$: when $\beta<\beta_s$, it is negative while
positive as $\beta >\beta_s$. This implies that a Hawking-Page
phase transition occurs when $\beta=\beta_s$. In the high
temperature phase, the black hole dominates and dual field theory
is in the deconfinement phase, while in the low temperature
phase, the AdS soliton dominates and dual field theory is in the
confined phase. The remarkable feature here is that the
deconfinement transition is completely determined by the ratio
$\beta/\beta_s$, the same as the case without the term ${\cal
O}(\alpha'^3)$, although the action gets corrected by the term
${\cal O}(\alpha'^3)$.

We find that the transition point keeps unchanged in the Witten's
QCD model can be understood from the point of view of the
Euclidean manifold $S^1\times S^1 \times R^2$, where the dual
field theory resides, where one of $S^1$ denotes the Euclidean
time $\tau$ with period $\beta$ (inverse temperature of the black
hole); the other stands for the compactified coordinate $x_1$ with
period $\beta_s$. The Hawking-Page transition happen when the
ratio $\beta/\beta_s$  decreases to one. When the $\gamma W$ term
is taken into account, we see from (\ref{2eq23}) and (\ref{3eq2})
that up to $ {\cal O}(\gamma^2)$, $\beta$ and $\beta_s$ depend on
$\gamma$ in a same manner. As a result, $\beta/\beta_s$ is
independent of $\gamma$; and therefore the deconfinement
transition is the same as the case without the correction term.
But, clearly both of $\beta$ and $\beta_s$ depend on the
correction term.

\sect{Conclusions and Discussions}

In this paper we have discussed Hawking-Page phase transitions
associated with Ricci flat $AdS_5$ black holes with correction
$\alpha'^3R^4$ in two situations. One is the so-called hard-wall
AdS/QCD model, where a small radius region of the AdS space is
removed by hand. Recently Herzog~\cite{Herzog} has shown that
deconfinement phase transition can be realized in the hard-wall
QCD model via a first-order Hawking-Page transition between a
Ricci flat $AdS_5$ black hole and the AdS space in the Poincare
coordinates. Removing the small radius region amounts to
introducing an IR cutoff in the dual field theory. Herzog has
found a relation between the deconfinement transition temperature
of the holographic QCD model and the IR cutoff. The $\alpha'$
expansion corrections in the gravity side are dual to the
$\lambda$ expansion corrections in the dual field theory with
$\lambda$ being the 't Hooft coupling constant. Therefore the
leading correction $\alpha'^3 R^4$ in the type IIB supergravity
gives the $\lambda^{-3/2}$ correction in the field theory side. We
have generalized the Herzog's discussion by studying the
$\alpha'^3R^4$ correction to the Ricci flat $AdS_5$ black holes
and found that the deconfinement temperature decreases after
considering the correction (see (\ref{2eq34})).

The other situation is that one of horizon coordinates is
compactified for the Ricci flat $AdS_5$ black hole. In fact this
situation is just the one for Witten's QCD model in $2+1$
dimensions. In this case, a suitable reference background is the
$AdS_5$ soliton solution, which has less energy than the pure
$AdS_5$ space. We have found that the deconfinement transition of
the QCD model is always determined by the ratio of inverse Hawking
temperature of the black hole to the period of the compactified
direction (see (\ref{3eq8})). The correction $\alpha'^3 R^4$ has
no effect on the ratio $\beta/\beta_s$. It would be very
interesting to see whether this conclusion is universal by
studying the deconfinement transition for Witten's QCD model in
$3+1$ dimensions. Very recently, the $\alpha'^3 R^4$ correction to
the $D4$-brane metric has been worked out by Basu~\cite{Basu}.

\section*{Acknowledgments}
The work of R.G.C. was supported in part by a grant from Chinese
Academy of Sciences, and by NSFC under grants No. 10325525 and No.
90403029.
The work of N.O. was supported in part by the Grant-in-Aid for
Scientific Research Fund of the JSPS Nos. 16540250 and 06042.

\end{document}